\documentclass{Interspeech2024}

\interspeechcameraready 

\usepackage{hyperref}
\usepackage{multirow}
\usepackage{multicol}
\usepackage{comment}
\usepackage{subfig}
\usepackage{amsmath}
\usepackage{amssymb}
\usepackage[vlined]{algorithm2e}
\usepackage{url}
\usepackage{xcolor}
\usepackage{float}
\usepackage{algpseudocode}
\usepackage{graphicx}
\usepackage{booktabs}
\usepackage{subfig}

\usepackage{subcaption}
\usepackage{soul}
\usepackage[table]{colortbl}

\title{AVR: Synergizing Foundation Models for Audio-Visual Humor Detection}
\name[affiliation={1}]{Sarthak}{Sharma*}
\name[affiliation={1}]{Orchid}{Chetia Phukan*}
\name[affiliation={1}]{Drishti}{Singh}
\name[affiliation={1}]{Arun}{Balaji Buduru}
\name[affiliation={1,2}]{Rajesh}{Sharma}

\address{
  $^1$IIIT-Delhi, India, 
  $^2$University of Tartu, Estonia \\ *equal contribution}
\email{sarthak23083@iiitd.ac.in , orchidp@iiitd.ac.in}

\keywords{Audio-Visual Humor Detection, Multimodal System, VideoMAE, Audio Spectrogram Transformer, Languagebind}

\begin{document}

\maketitle

\begin{abstract}
    
\noindent In this work, we present, \textbf{AVR} application for audio-visual humor detection. While humor detection has traditionally centered around textual analysis, recent advancements have spotlighted multimodal approaches. However, these methods lean on textual cues as a modality, necessitating the use of ASR systems for transcribing the audio-data. This heavy reliance on ASR accuracy can pose challenges in real-world applications. To address this bottleneck, we propose an innovative audio-visual humor detection system that circumvents textual reliance, eliminating the need for ASR models. Instead, the proposed approach hinges on the intricate interplay between audio and visual content for effective humor detection.
\end{abstract}

\section{Introduction}


Humor has intrigued researchers since ancient Greek times \cite{larkin2017overview}, and numerous theories have been proposed since then. Theories have suggested that humor encompasses a combination of semantic context, facial expressions, gestures, and voice modulation, rendering it complex for even humans to detect in various situations \cite{weber2018shape}. These challenges have spurred researchers to delve into humor detection, making it a crucial domain in human-computer interaction (HCI).

The exploration of humor detection commenced with the analysis of textual content around 2004 \cite{taylor2004computationally}, primarily focusing on jokes and puns characterized by word-play and categorizing such sentences as humorous or not. Initially, NLP techniques such as N-grams were employed to assess the similarity scores for predicting the humor quotient of sentences. Subsequently, the scope of datasets expanded to include data from diverse sources like social media platforms such as Twitter \cite{potash2017semeval} and Reddit. Machine learning models such as Support Vector Machine (SVM) and deep learning models such as CNNs and LSTMs were utilized for prediction purposes. To further enhance humor detection, modalities like audio and video were introduced to incorporate facial expressions and voice modulations \cite{weller2019humor}. TV sitcoms such as "The Big Bang Theory" \cite{bertero2016deep} and Friends have been frequently employed as datasets for multimodal humor analysis due to their incorporation of various humor types such as sarcasm, puns, and satire. Another study \cite{christ2022towards} utilized the novel Passau-Spontaneous Football Coach Humour (Passau-SFCH) dataset, and trained the models on each modality - speech, text and video and got promising results for the multimodal approach. 

The reliance on textual cues for multimodal humor detection systems, which often require accurate transcription through ASR systems, can indeed present challenges, particularly in real-world applications where ASR accuracy may vary. To overcome this, we exclusively fuse audio-visual information for more reliable humor detection, removing the need for ASR models. Leveraging state-of-the-art foundation models like VideoMAE, AST, and Languagebind, we extract meaningful audio-visual features for effective humor detection and we name our application as \textbf{AVR}. \textbf{AVR} intakes a whole video and provides the output without the need for analyzing individual frames in a video.

\begin{figure}
\centering
\includegraphics[width=0.40\textwidth, height=0.45\textwidth]{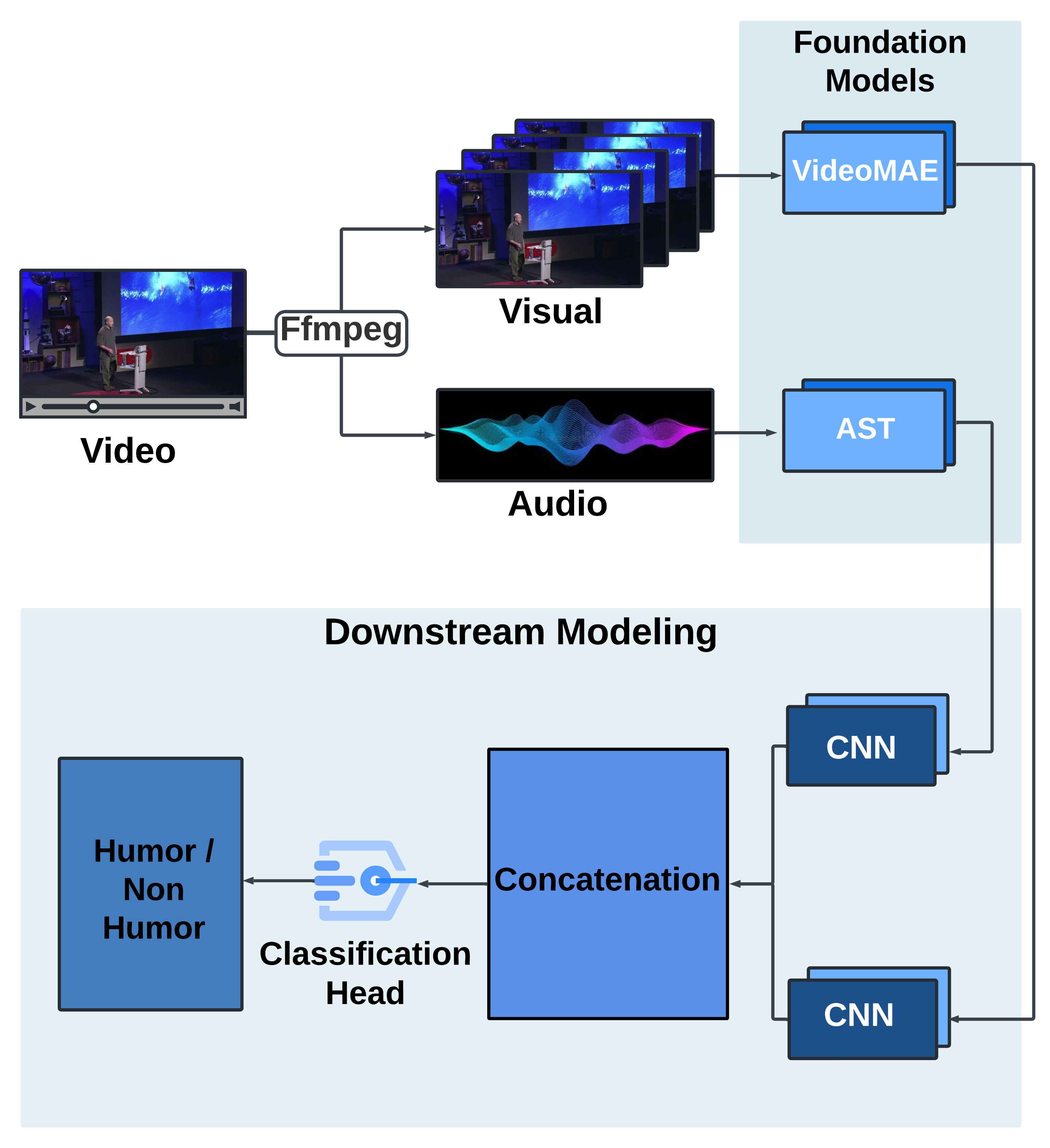}
\caption{
Architecture of the Proposed Modeling Network
}
\label{fig:obsssn}
\end{figure}

\section{AVR}
In this section, we delve into various parts of \textbf{AVR} application. First, the foundation models, followed by the downstream modeling, experimental analysis, and lastly the user interface. 

\noindent\textbf{Foundation Models}: We use Languagebind \cite{zhu2023languagebind}, which is a multimodal model trained for multiple modalities with language as the binding modality. Languagebind shows state-of-the-art (SOTA) performance in various tasks across different modalities. We also use VideoMAE \cite{tong2022videomae}, AST \cite{gong21b_interspeech} which are foundation models trained for video, audio representation learning in self-supervised and supervised fashion. We extract audio-visual representations of 768-dimension from the foundation models. The audios are resampled to 16KHz before passing it to the foundation models.

\begin{table}
\centering
\setlength{\tabcolsep}{15pt}
\caption{Evaluation Scores; Scores are average of 5-folds; F1 stands macro-average f1-score}
\label{tab:results}
\begin{tabular}{lc} \toprule
\textbf{\textcolor{blue}{Foundation Models}} & \textbf{\textcolor{blue}{Accuracy}} \\ \midrule
\multicolumn{2}{c}{\textbf{\textcolor{red}{CNN}}} \\ \midrule
LanguageBind          &     49.68 \\
VideoMAE + AST        &    \textbf{56.70}        \\
 \midrule  
\multicolumn{2}{c}{\textbf{\textcolor{red}{LSTM}}} \\ \midrule
    LanguageBind               &     48.40          \\
VideoMAE + AST                 &     54.82    \\
\bottomrule
\label{scores}
\end{tabular}
\end{table}

\noindent\textbf{Downstream Modelling}: We have employed CNN and LSTM as downstream networks. The modeling architecture with CNN as downstream is shown in Figure \ref{fig:obsssn}. The CNN model consists of two 1D-CNN with 32 and 64 filters and 3 as filter size. Followed by concatenation for fusion and lastly, classification head with a hidden layer of 128 neurons followed by the output layer. For LSTM model, we use a single LSTM layer with 50 as intermediate followed by the same classification head and the corresponding layer. The models are trained for 50 epochs with a learning rate of 1e-5. We also employed dropout and early stopping forreducing overfitting. We train the models in a 5-fold manner with 4 folds as training set and one fold as test set. We use cross-entropy as loss function and Adam as the optimizer. The results of the models are given in Table \ref{scores}. We can see that CNN with VideoMAE and AST representations earned the topmost score. One notable thing is the performance of VideoMAE and AST representations compared to LanguageBind despite it being jointly trained on multiple modalities.

\begin{figure}
\centering
\includegraphics[width=0.4\textwidth, height=0.5\textwidth]{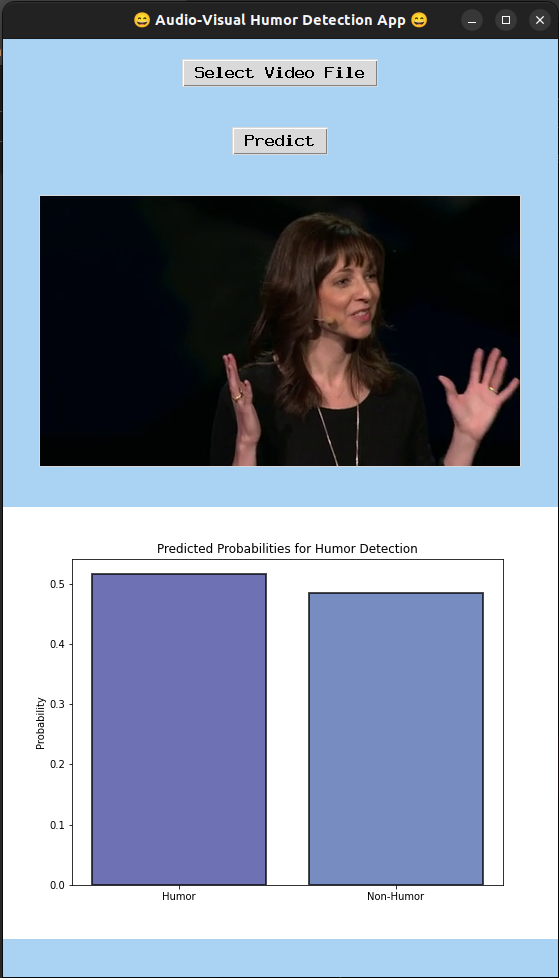}
\caption{
User Interface
}
\label{ui}
\end{figure}

\subsection{User Interface}
To facilitate efficient user engagement with the model, we have developed an interactive User Interface (UI) leveraging \textit{Tkinter} library. \textbf{AVR} provides the benefit of cross-system usage across macOS, Linux, windows, and so on. The UI of \textbf{AVR} is shown in Figure \ref{ui}. Users are prompted to select a .mp4 video file from their local device to upload it. On uploading, the video is segmented into audio and visual modality using \textit{FFmpeg}. Similar to the model training, representations are generated for both the audio and video using AST and VideoMAE respectively and passed to CNN (we use CNN in backend of \textbf{AVR} as it shows the best performance).  
Users are presented with the predictions with a graphical representation in the form of a bar graph, illustrating the probability of the input video being categorized as humor or non-humor. \textbf{AVR} takes approx 5-10 secs for inference for videos of duration of 2-30 secs in a P100 GPU. 

\section{Conclusion}
We present \textbf{AVR}, an innovative application for audio-visual humor detection using foundation models. This synergy of audio and visual modalities for effective humor detection and thus prevents the usage of text as a modality in multimodal humor detection and in return reducing the dependency on ASR system accuracy. 


\bibliographystyle{IEEEtran}
\bibliography{main.bib}

\end{document}